\documentclass[aps,floatfix,nofootinbib,onecolumn,superscriptaddress]{revtex4}
\pdfoutput=1

\usepackage{amsmath}
\usepackage{amssymb}
\usepackage{amsthm}
\usepackage{dcolumn}
\usepackage{epsfig}
\usepackage{graphics}
\usepackage{graphicx}
\usepackage{slashed,epsfig}
\usepackage{longtable}
\usepackage{color}

\definecolor{darkgreen}{rgb}{0,0.5,0}
\definecolor{purple}{rgb}{0.5,0,0.5}
\definecolor{nblue}{rgb}{0.0,0.0,0.50}
\definecolor{scarlet}{rgb}{1.0,0.2,0}
\usepackage[colorlinks=true, pdfstartview=FitV, linkcolor=purple, citecolor= purple, urlcolor=blue]{hyperref}

\newcommand{\beq} {\begin{equation}}
\newcommand{\eeq} {\end{equation}}
\newcommand{\beqa} {\begin{eqnarray}}
\newcommand{\eeqa} {\end{eqnarray}}

\begin{document}

{\par\raggedleft \texttt{SLAC-PUB-14757}\par}
\bigskip{}

\title{Atoms in Flight and the  Remarkable Connections between
Atomic and Hadronic Physics}

\author{Stanley~J.~Brodsky} \affiliation{SLAC National Accelerator Laboratory\\
Stanford University, Stanford, California 94309, USA}

\begin{abstract}

Atomic physics and hadron physics are both based on  Yang Mills gauge theory; in fact, quantum electrodynamics  can be regarded as the zero-color limit of quantum chromodynamics. 
I review a number of areas where the techniques of atomic physics provide important insight into the theory of hadrons in QCD.  For example, the Dirac-Coulomb equation, which predicts the spectroscopy and structure of hydrogenic atoms, has an analog in hadron physics in the form of light-front relativistic equations of motion which give a remarkable first approximation to the spectroscopy, dynamics, and structure of light hadrons. The renormalization scale for the running coupling, which is unambiguously set in QED, leads to a method for setting the renormalization scale in QCD. The production of atoms in flight provides a method for computing the formation of hadrons at the amplitude level.  Conversely, many techniques which have been developed for hadron physics, such as scaling laws, evolution equations, and light-front quantization have equal utility for atomic physics, especially in the relativistic domain.
I also present a new perspective for understanding the contributions to the cosmological constant from QED and QCD.

\end{abstract}


\maketitle

\date{\today}

\section{Introduction}
\label{intro}
Quantum Electrodynamics, the fundamental theory of leptons and photons which underlies all of atomic and molecular physics,  and Quantum Chromodynamics,  the quark and gluon theory  with three colors  underlying hadronic and nuclear physics,   are both derived from Yang-Mills gauge theory. The Yang-Mills Lagrangian for $SU(N_C)$  is invariant under arbitrary color rotations and phases  at each point of space and time.   In fact, in the limit  where the number of colors $N_C$ vanishes,  with $\alpha_s C_F = \alpha$ held fixed ($C_F \equiv (N^2_C- 1) /2 N_C$),  QCD becomes equivalent to Abelian gauge theory~\cite{Brodsky:1997jk}.  This analytic connection as a function of $N_C$  between QCD and QED  provides a valuable link between the two fields;   processes  and analyses in QCD must connect at zero color to the analogous reactions and procedures of QED.

In this paper, I will review a number of areas where the techniques of atomic physics give important insight into the theory of hadrons, the color-singlet bound states of quarks and gluons in QCD.  For example, the Dirac-Coulomb equation, which predicts the spectroscopy and structure of hydrogenic atoms has an analog in hadron physics in the form of relativistic frame-independent equations of motion derived from light-front holography~\cite{deTeramond:2008ht} which give a remarkable first approximation to the spectroscopy, dynamics, and structure of light hadrons. The renormalization scale for the running coupling which is unambiguously set in QED leads to a solution for setting the renormalization scale in QCD. The production of atoms in flight provides a method for computing the formation of hadrons at the amplitude level. Conversely, many techniques and theorems developed for hadron physics, such as scaling laws, evolution equations, and light-front quantization have equal utility for atomic physics, especially in the relativistic domain.

\section{Production of Exotic Atoms in Flight and Hadronization at the Amplitude Level }
\begin{figure}
 \begin{center}
\includegraphics[width=16cm]{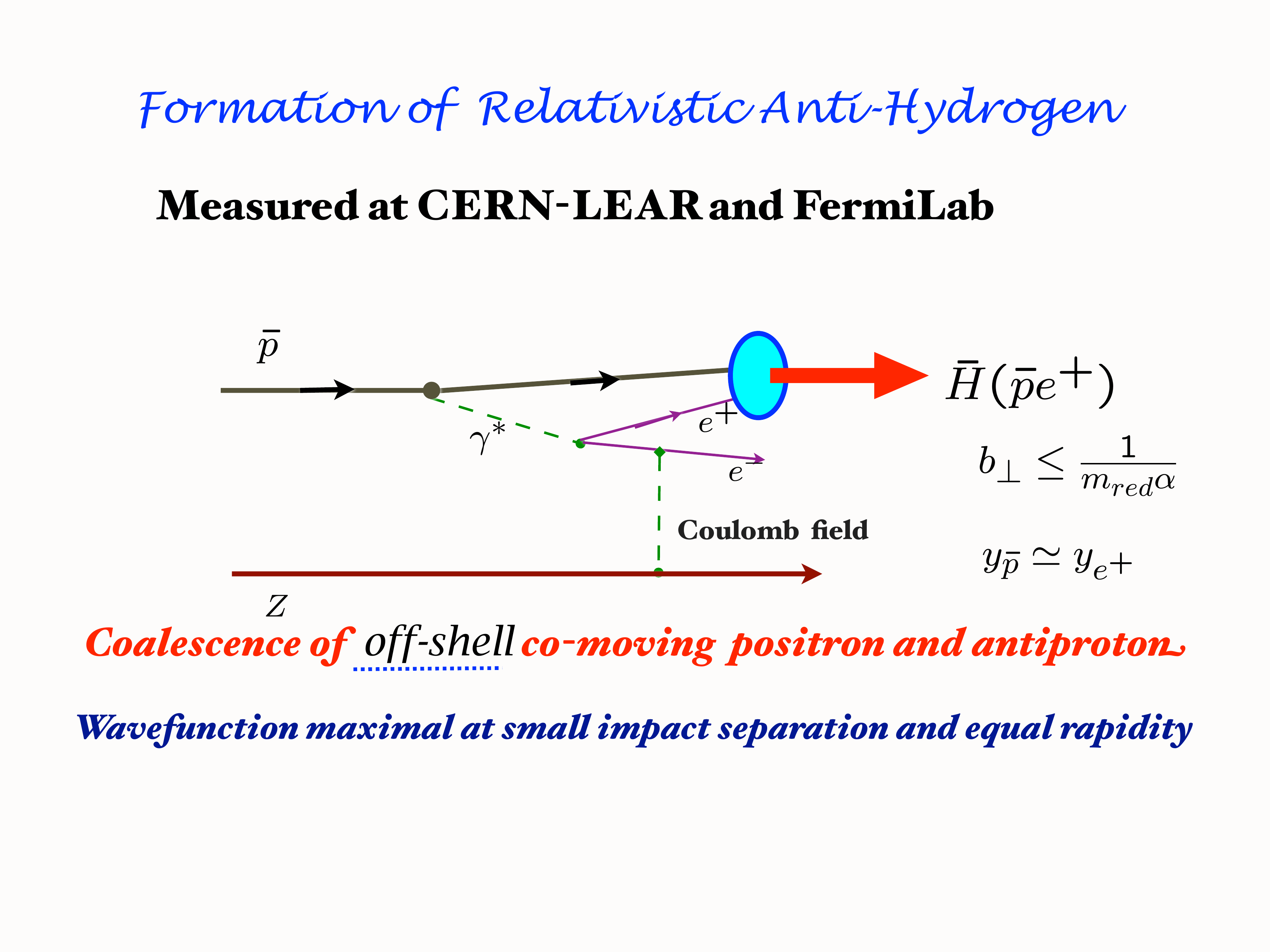} 
\includegraphics[width= 16cm]{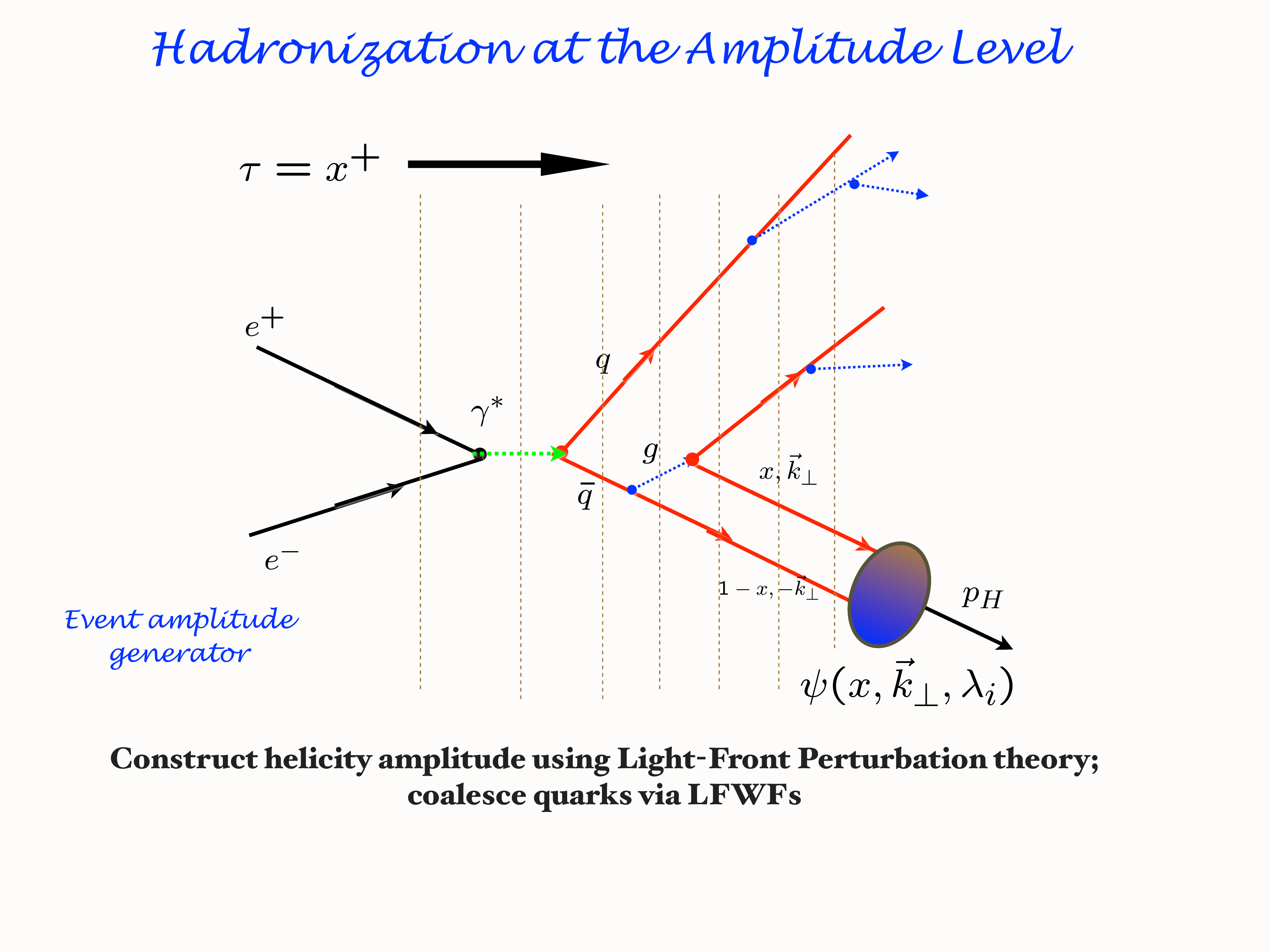}
\end{center}
\caption{(a) Production of Relativistic Antihydrogen. (b) Hadronization at the Amplitude Level in Electron-Positron Annihilation}
\label{Antihydrogen}  
\end{figure}

Relativistic antihydrogen was first produced in 1995 at CERN-LEAR~\cite{Baur:1995ck} and at the Fermilab Antiproton Accumulator~\cite{Blanford:1997up}. The production mechanism~\cite{Munger:1993kq} is illustrated in fig. \ref{Antihydrogen} (a). The incident antiproton beam produces a Bethe-Heitler electron-positron pair in the Coulomb field of a target nucleus $ \bar p Z \to \bar p e^+ e^- Z  \to [\bar p e^+] Z$. The comoving off-shell $\bar p$ and $e^+$ then coalesece into antihydrogen atoms via the Schr\"odinger Coulomb wavefunction which connects the off-shell state to the on-shell anti-atom. The atom is dominantly in its 1S ground state. In principle, one can measure  its ``anti-Lamb-Shift" using  the Robiscoe level-crossing method~\cite{Robiscoe:1965zz}.  

The production of antihydrogen in flight provides important insight into the dynamics of hadron production in QCD.  
For example,  the  $\Lambda(sud)$ baryon can be produced at high longitudinal momentum fraction $x_F$ in $ p p \to \Lambda X$ reactions by the coalescence of the $ud$ valence quarks of the beam with a comoving strangeness quark. This method can be generalized to produce heavy hadrons such as $\Lambda_c(cud), \Lambda_b,$  double charmed baryons, etc., using the high $ x$ intrinsic heavy quarks which exist in the higher Fock states of  the proton wavefunction~\cite{Brodsky:1981se}.

The analog of intrinsic charm in hadrons is the $\mu^+ \mu^-$ content of positronium. The $|e^+ e^- \mu^+ \mu^->$ Fock state appears through the cut of the muon-loop light-by-light contribution to the self energy of the positronium eigenstate.   In this Fock state, the muons carry almost all of the momentum of the moving atom since the off-shell virtuality is minimal at equal velocity.  In QED the probability for intrinsic leptons $L \bar L$  exist in positronium scales as $1/m^4_L$ whereas in QCD the probability of intrinsic heavy quarks in the wavefunction of a light hadron scales as  $1/ m^2_Q$ because of its non-Abelian couplings~\cite{Brodsky:1984nx,Franz:2000ee}.
 
The production of a $q \bar q$ meson in an $e^+ e^-$ annihilation event is illustrated in fig. \ref{Antihydrogen}(b).  One first calculates the $T$ matrix element for the production of off-shell quarks and gluons at the amplitude level using light-front time-ordered perturbation theory.  The light-front wavefunction of the meson then converts the off-shell comoving $q \bar q$ pair into the final-state meson.   The confined colored quarks thus never appear on-shell. This first-principle method for forming hadrons in QCD~\cite{Brodsky:2008tk} can replace phenomenological jet hadronization models such as PYTHIA.
The  light-front wavefunction required for calculating ``hadronization at the amplitude level" ~\cite{Brodsky:2008tk,Brodsky:2009dr}is the frame-independent analog of the Schr\"odinger wavefunction of atomic physics. It is obtained from the eigensolution of the QCD light-front Hamiltonian quantized at fixed light-front time $\tau$   which can be determined by solving the Heisenberg matrix $H_{LF}^{QCD}|\Psi_H> = M^2_H |\Psi_H>$ using a method such as discretized light-cone quantization (DLCQ)~\cite{Pauli:1985pv} or using the AdS/QCD approach together with Light-Front Holography~\cite{deTeramond:2008ht}.

\begin{figure}
 \begin{center}
\includegraphics[width=18cm]{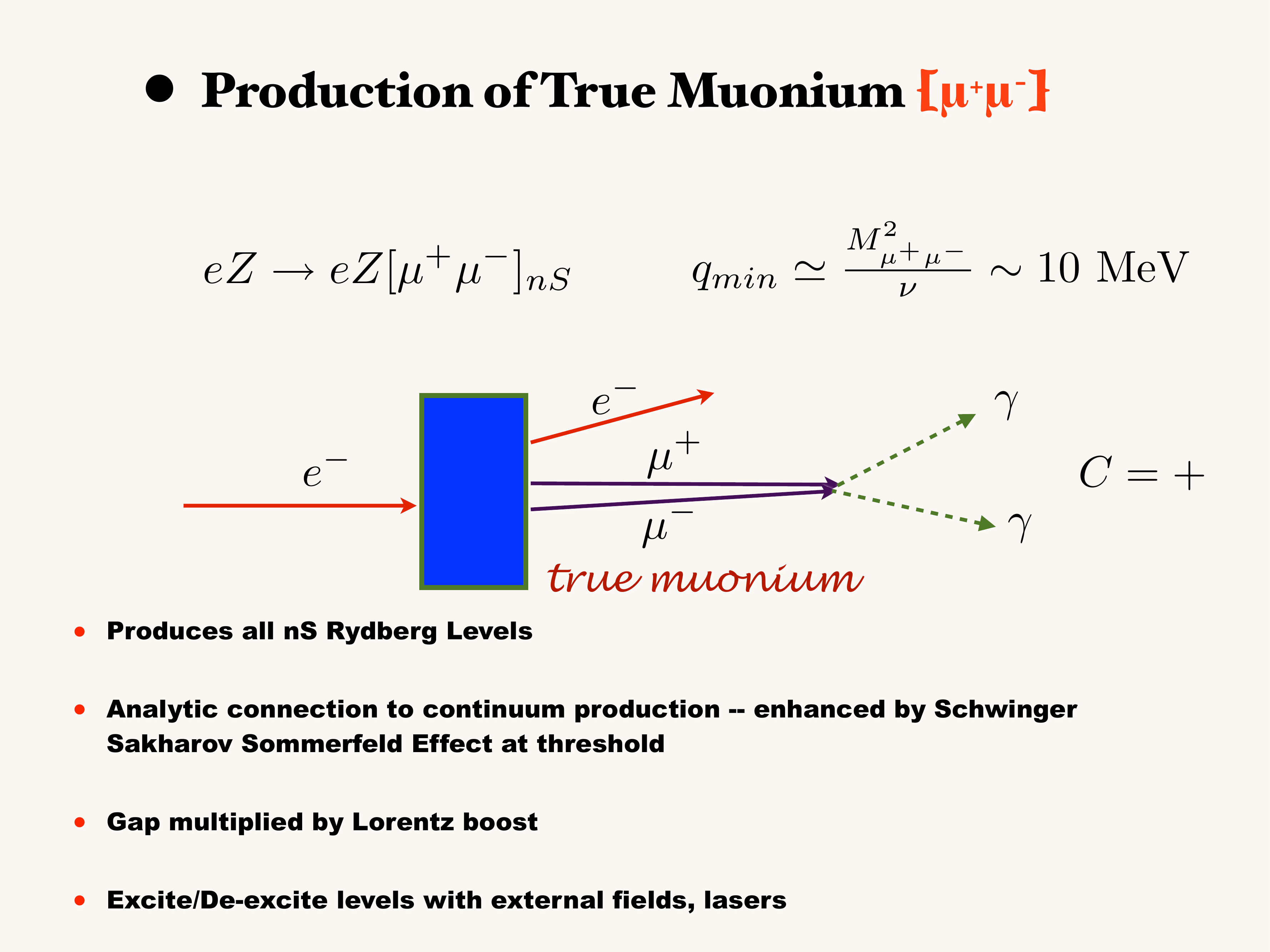}
\end{center}
\caption{ Production of True Muonium}
\label{True}  
\end{figure}

It is very interesting to produce  ``true muonium",  the  $[\mu^+ \mu^-]$ bound state  
Lebed and I~\cite{Brodsky:2009gx} have discussed QED production and decay mechanisms, such as electroproduction of relativistic true muonium
below the $\mu^+ \mu^-$  threshold via 
$e^- Z \to [\mu^+ \mu^-]  e^- Z$  or $e^+ e^- \to [\mu^+ \mu^-] \gamma$.  See fig. \ref{True}.   The APEX  electroproduction experiment~\cite{Abrahamyan:2011gv}, which will search for dark matter candidates at Jefferson Laboratory, could be the first to see this exotic atom. 
Studying the precision spectroscopy of the $[\mu^+ \mu^-]$
atom is important in view of the anomalies seen in the muon $g-2$~\cite{Bennett:2006fi} and the $\mu^- p$ Lamb shift~\cite{Pohl:2010zz}.

``Atomic Alchemy "refers to the transition between a muonic atom into an electronic atom: $(\mu^- Z) \to (e^- Z) {\bar \nu_e }\nu_\mu $ via the weak decay of the bound muon and the subsequent capture of its decay electron.    Greub, Wyler, Munger and I ~\cite{Greub:1994fp} have shown that such processes provide a laboratory for studying the relativistic high momentum tail of wavefunctions in atomic physics; in addition, they provide a simple toy model for investigating analogous exclusive heavy hadronic decays in quantum chromodynamics such as $B \to \pi e \nu.$

The  QCD analog of a molecule in QCD is a  bound state of heavy quarkonium with a nucleus such as $[J/\psi A]$~\cite{Brodsky:1989jd,Luke:1992tm}. The binding occurs through two-gluon exchange,  the hadronic analog of the Van der Waals interaction. Since the kinetic energy of the $J/\psi$ and the nucleus are both small, one expects to find  produce these exotic hybrid states at threshold. Examples of nuclear-bound quarkonium are  the $|uud uud s \bar s>$ and $|uud uud c \bar c>$ resonances which apparently appear as intermediate states in $p p \to p p $ elastic exchange. These resonances can account~\cite{Brodsky:1987xw} for the large spin-spin $A_{NN}$ correlations~\cite{Court:1986dh} observed at the strangeness $E_{cm} \simeq 3 $ GeV and $E_{cm} \simeq 5$ GeV and charm thresholds.

At high energies,  Compton scattering on an atom $\gamma A \to \gamma A$  is dominated by the Thomson amplitude -- the elastic scattering of the photon on the atomic electrons.  The analog in hadron physics is the scattering of photons on quarks $\gamma q \to \gamma q$ via a local seagull or instantaneous light-front term or which gives an energy-independent contribution to the Compton amplitude proportional to the charge squared of the struck quark --  a contribution which has no analog in hadron scattering reactions.  Llanes-Estrada, Szczepaniak, and I~\cite{Brodsky:2008qu} have shown that this local contribution has a real phase and is universal, giving the same contribution for real or virtual Compton scattering for any photon virtuality and skewness at fixed momentum transfer squared $ t$. The $t$-dependence of this $J=0$  fixed Regge pole is parameterized by a yet unmeasured even charge-conjugation form factor of the target nucleon. The $t=0$ limit gives an important constraint on the dependence of the nucleon mass on the quark mass through the Weisberger relation.  The same  $J=0$ amplitude enters the two-photon exchange contribution to muon-proton scattering, and thus also could contribute an important contribution to the $\mu p$ Lamb Shift.

\section{Renormalization Scale setting}
A key difficulty in making precise perturbative predictions for QCD is
the uncertainty in determining the renormalization scale $\mu$ of the
running coupling $\alpha_s(\mu^2)$.  In the standard Gell-Mann--Low
scheme for QED, the renormalization scale is simply the virtuality
of the virtual photon~\cite{GellMann:1954fq}. 
Although the {\it initial} choice of renormalization scale
$t_0$ is arbitrary, the {\it final} scale $t$ which  sums the
vacuum polarization corrections is unique and unambiguous. The
resulting perturbative series is identical to the conformal series
with zero $\beta$-function. In the case of muonic atoms, the
modified muon-nucleus Coulomb potential is precisely
$-Z\alpha(-{\vec q}^{~2})/ {\vec q}^{~2};  $  i.e., $\mu^2=-{\vec
q}^2.$ Again, the renormalization scale is unique.
The same principle underlying renormalization scale-setting in QED for $N_C=0$
must also hold in QCD since the $n_F$ terms in the QCD $\beta$ function
have the same role as the lepton $N_\ell$ vacuum polarization
contributions in QED.  
Thus the same scale-setting procedure must be applicable to all renormalizable gauge theories.

The purpose of the running coupling in any gauge theory is to sum all
terms involving the $\beta$ function; in fact, when the
renormalization scale $\mu$ is set properly, all non-conformal
$\beta \ne 0$ terms  in a perturbative expansion arising from
renormalization are summed into the running coupling.  The
remaining terms in the perturbative series are then identical to
that of a conformal theory; i.e., the theory with $\beta=0$.  The
divergent ``renormalon" series of order $\alpha_s^n \beta^n n! $
does not appear in the conformal series. Thus  as in quantum electrodynamics, the
renormalization scale $\mu$ is determined unambiguously by the
``Principle of Maximal Conformality (PMC)" ~\cite{Brodsky:2011ig,Brodsky:2011ta}.  This is also the principle
underlying BLM scale setting~\cite{Brodsky:1982gc}
An important
feature of the PMC is that its QCD predictions are independent of
the choice of renormalization scheme. The PMC procedure also
agrees with QED scale-setting in the $N_C \to 0$ limit.

 \section{Light-Front Quantization}

The distributions of electrons within an atom are determined in QED using the Schr\"odinger wavefunction, the eigenfunction of the QED Hamiltonian.  In principle, one could  calculate hadronic spectroscopy and wavefunctions by solving for the eigenstates of the QCD Hamiltonian: 
$H \vert  \Psi \rangle = E \vert \Psi \rangle$
 at fixed time $t.$ However, this traditional method -- called the ``instant" form" by Dirac,~\cite{Dirac:1949cp} is  plagued by complex vacuum and relativistic effects, as well 
as  by  the fact that the boost of such fixed-$t$ wavefunctions away from the hadron's rest frame is an intractable dynamical problem.  
However, there is an extraordinarily powerful non-perturbative alternative -- quantization at fixed light-front (LF) time $\tau = t + z/c = x^+ = x^0 + x^3$ -- the ``front-form" of Dirac.~\cite{Dirac:1949cp} In this framework each hadron $H$ is identified as an eigenstate of the QCD Hamiltonian 
$H_{LF}^{QCD} \vert \Psi_H \rangle = M^2_H \vert \Psi_H \rangle$,   
where  $H_{LF}^{QCD} = P_\mu P^\mu= P^- P^+ -  P^2_\perp$ is derived directly from the QCD Lagrangian or action. The eigenvalues of this Heisenberg equation give the complete mass spectrum of hadrons. The eigensolution  $|\Psi_H \rangle$  projected on the free Fock basis  provides  the  set of valence and non-valence  light-front Fock state wavefunctions $\Psi_{n/H}(x_i, k_{\perp i}, \lambda_i)$, which describe the hadron's momentum and spin distributions and  the direct measures of its structure at the quark and gluon level.   
If one quantizes the gluon field in light-cone gauge $A^+= A^0 + A^3=0$, the gluons have physical polarization $S^z = \pm 1$, there are no ghosts, so that  one has a physical interpretation of the quark and gluon constituents. 
The constituents of a bound state in a light-front wavefunction are measured at the same light-front time $\tau$ -- along the front of a light-wave, as in a flash picture.  In contrast, the constituents of a bound state in an instant form wavefunction must be measured at the same instant time $t$ -  - this requires the exact synchrony in time of many simultaneous probes. 

A remarkable feature of LFWFs is the fact that they are frame
independent; i.e., the form of the LFWF is independent of the
hadron's total momentum $P^+ = P^0 + P^3$ and $P_\perp.$
The boost invariance of  LFWFs contrasts dramatically with the complexity of  boosting the wavefunctions defined at fixed time $t.$~\cite{Brodsky:1968ea}  
Light-front quantization is thus the ideal framework to describe the
structure of hadrons in terms of their quark and gluon degrees of freedom.  The
constituent spin and orbital angular momentum properties of the
hadrons are also encoded in the LFWFs.  
The total  angular momentum projection~\cite{Brodsky:2000ii} 
$J^z = \sum_{i=1}^n  S^z_i + \sum_{i=1}^{n-1} L^z_i$ 
is conserved Fock-state by Fock-state and by every interaction in the LF Hamiltonian.
The constituent spin and orbital angular momentum properties of the hadrons are thus encoded in their LFWFs.   
The empirical observation that quarks carry only a small fraction of the nucleon angular momentum highlights the importance of quark orbital angular momentum.  In fact the nucleon anomalous moment and the Pauli form factor are zero unless the quarks carry nonzero $L^z$.

Hadron observables, e.g., hadronic structure functions, form factors, distribution amplitudes,  GPDs, TMDs, and Wigner distributions can be computed as simple convolutions of light-front wavefunctions (LFWFs).  For example,  one can calculate the electromagnetic and gravitational form factors 
$<p+ q| j^\mu(0)| p>$ and $<p+ q| t^{\mu \nu}(0)| p>$ of a hadron from the Drell-Yan-West formula -- i.e., the overlap of LFWFs.    
The anomalous gravitomagnetic moment $B(0)$ defined from the spin-flip matrix element  
$<p+ q| t^{\mu \nu}(0)| p>$ at $ q\to 0$ vanishes -- consistent with the equivalence theorem of gravity.
In contrast, in the instant form, the overlap of instant time wavefunctions is not sufficient.  One must also couple the photon probe to currents arising spontaneously from the vacuum which are connected to the hadron's constituents. 
The Light-Front method is  directly applicable for describing atomic bound states in both the relativistic and nonrelativistic domains; it is particularly useful for atoms in flight since the LFWFs are frame-independent. It also satisfies theorems 
such as cluster decomposition.

One can solve the LF Hamiltonian problem for theories  in one-space and one-time  by Heisenberg matrix diagonalization. For example, the complete set of discrete and continuum eigensolutions of mesons and baryons  in QCD(1+1) can be obtained to any desired precision for general color,  multiple flavors, and general quark masses using the discretized light-cone quantized (DLCQ) method.~\cite{Pauli:1985ps,Hornbostel:1988fb}  The  DLCQ approach can in principle be applied to QED(3+1) and QCD(3+1); however,  in practice, the  huge matrix diagonalization problem is computational challenging.

\section{AdS/QCD Light-Front Holography}
Recently a new nonperturbative QCD  approach has been developed which leads to an elegant analytical
 and phenomenologically compelling first approximation to the full LF Hamiltonian method-- ``Light-Front Holography".~\cite{deTeramond:2008ht}
Light front holographic methods
allow one to project the functional dependence of the wavefunction $\Phi(z)$ computed  in the  AdS fifth dimension to the  hadronic frame-independent light-front wavefunction $\psi(x_i, b_{\perp i})$ in $3+1$ physical space-time. The variable $z $ maps  to a transverse
LF variable $ \zeta(x_i, b_{\perp i})$. 
The result is a single-variable light-front Schr\"odinger equation which determines the eigenspectrum and the LFWFs of hadrons for general spin and orbital angular momentum.  The transverse coordinate $\zeta$ is closely related to the invariant mass squared  of the constituents in the LFWF  and its off-shellness  in  the LF kinetic energy,  and it is thus the natural variable to characterize the hadronic wavefunction.  In fact $\zeta$ is the only variable to appear 
in the relativistic light-front Schr\"odinger equations predicted from 
holographic QCD  in the limit of zero quark masses. 
The coordinate $z$ in AdS space is thus uniquely identified with  a Lorentz-invariant  coordinate $\zeta$ which measures the separation of the constituents within a hadron at equal light-front time. 

The hadron eigenstates generally have components with different orbital angular momentum; e.g.,  the proton eigenstate in LF holographic QCD  with massless quarks has $L=0$ and $L=1$ light-front Fock components with equal probability.   Higher Fock states with extra quark-anti quark pairs also arise.   The resulting LFWFs then lead to a new range of hadron phenomenology,  including the possibility to compute the hadronization of quark and gluon jets at the amplitude level. The soft-wall model also predicts the form of the non-perturbative effective coupling and its $\beta$-function.~\cite{Brodsky:2010ur}  

\section{Lensing and the Sivers Effect}

A well-known phenomenon in QED  rescattering via final-state Coulomb interactions. Although the Coulomb phase for a given partial wave is infinite, the interference of Coulomb phases arising from different partial waves leads to observable effects. 

\begin{figure}
 \begin{center}
\includegraphics[width=16cm]{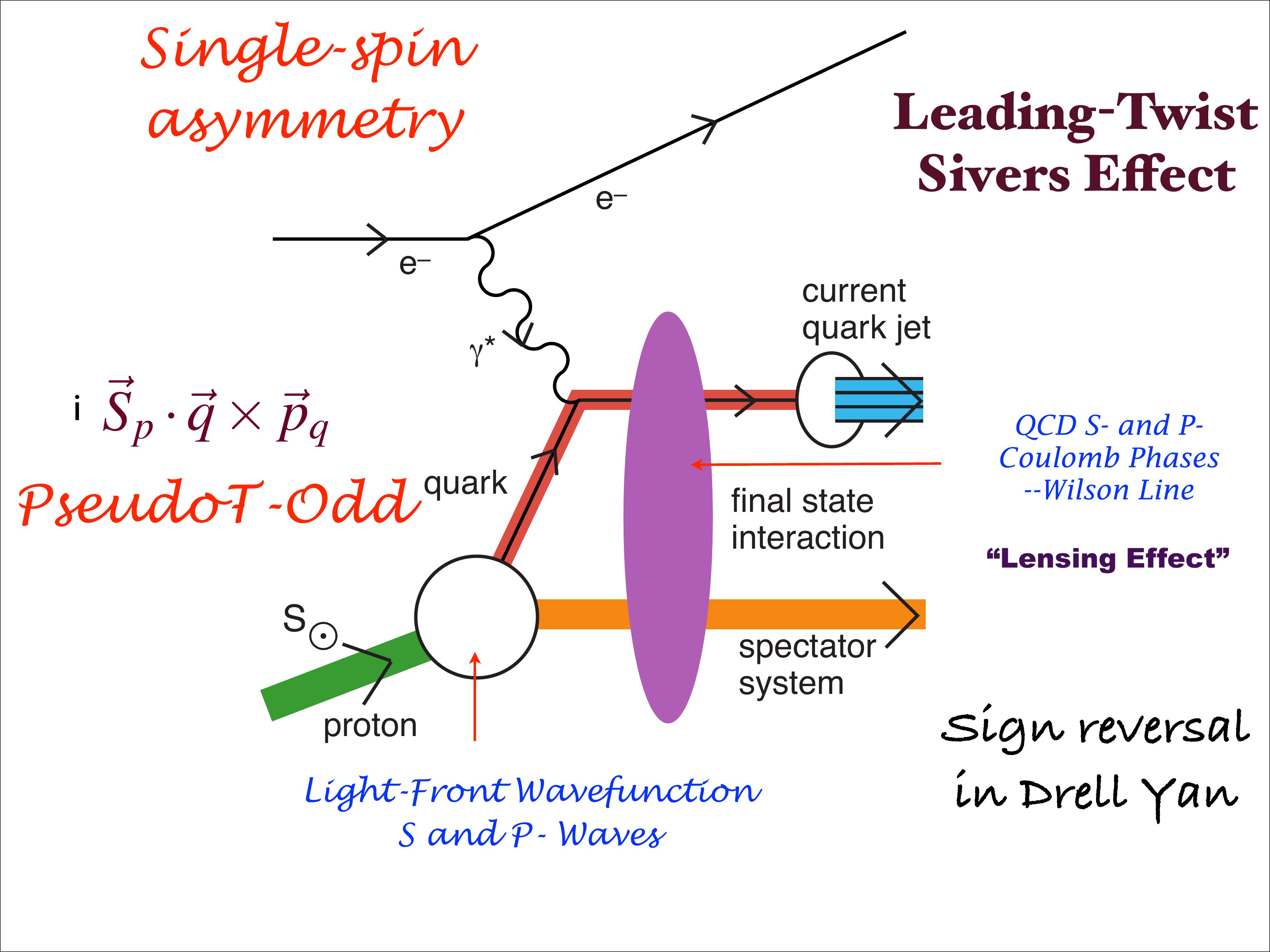}
\end{center}
\caption{Origin of the Sivers single-spin asymmetry in deep inelastic lepton scattering.}
\label{Sivers}  
\end{figure}

The calculation of the Sivers single-spin asymmetry in deep inelastic lepton scattering in QCD is illustrated in fig.~\ref{Sivers}.
The analysis requires two different orbital angular momentum components: $S$-wave with the quark-spin parallel to the proton spin and $P$-wave for the quark with anti-parallel spin; the difference between the final-state ``Coulomb" phases leads to a $\vec S \cdot \vec q \times \vec p$ correlation of the proton's spin with the virtual photon-to-quark production plane~\cite{Brodsky:2002cx}.  Thus, as it is clear from its QED analog,  the final-state gluonic interactions of the scattered quark lead to a  $T$-odd non-zero spin correlation of the plane of the lepton-quark scattering plane with the polarization of the target proton~\cite{Brodsky:2002cx}.  This  leading-twist Bjorken-scaling ``Sivers effect"  is nonuniversal since QCD predicts an opposite-sign correlation~\cite{Collins:2002kn,Brodsky:2002rv} in Drell-Yan reactions due to the initial-state interactions of the annihilating antiquark. 
The  $S-$ and $P$-wave proton wavefunctions also appear in the calculation of the Pauli form factor quark-by-quark. Thus one can correlate the Sivers asymmetry for each struck quark with the anomalous magnetic moment of the proton carried by that quark~\cite{Lu:2006kt},  leading to the prediction that the Sivers effect is larger for  positive pions.
The physics of the lensing dynamics involves nonperturbative quark-quark interactions at small momentum transfer, not  the hard scale $Q^2$  of the virtuality of the photon.  It would interesting to see if the strength of the 
soft initial- or final- state scattering can be predicted using the confining potential of AdS/QCD.

\section{Vacuum Condensates and the Cosmological Constant}

It is important to distinguish two very different concepts of the vacuum in quantum field theories such as QED and QCD.
The vacuum is normally defined as the lowest energy eigenstate of the instant-form Hamiltonian --  the vacuum defined by quantizing at fixed time $t$.  In QED, the instant-time vacuum is saturated with quantum loops of leptons and photons. In calculations of physical processes one must normal-order the vacuum and divide the $S$-matrix elements by the disconnected vacuum loops.
In contrast, the front-form (light-front) vacuum is defined as the lowest mass eigenstate of light-front Hamiltonian defined by quantizing at fixed $\tau = t -z/c$. The  vacuum is remarkably simple in light-front quantization because of the restriction $k^+ \ge 0.$   For example QED vacuum graphs such as $e^+ e^- \gamma $ do not arise.   The LF vacuum thus coincides with the  vacuum of the free LF Hamiltonian.  The front-form vacuum and  its eigenstates are Lorentz invariant; whereas the instant form vacuum depends on the observer's Lorentz frame. 
The instant-form vacuum is a state defined at the same time $t$ at all spatial points in the universe.  In contrast, the front-from vacuum only senses phenomena which are causally connected; i.e., or within the observer's light-cone.
Causality in quantum field theory follows the fact that commutators vanish outside the light-cone. In fact in the LF analysis  the spatial support of QCD condensates
is restricted to the interior of hadrons, physics which arises due to the
interactions of confined quarks and gluons.  The condensate physics is replaced by the dynamics of higher non-valence Fock states as shown by Casher and Susskind.~\cite{Casher:1974xd}  In particular, chiral symmetry is broken in a limited domain of size $1/ m_\pi$,  in analogy to the limited physical extent of superconductor phases.  
This novel description  of chiral symmetry breaking  in terms of ``in-hadron condensates"  has also been observed in Bethe-Salpeter studies~\cite{Maris:1997hd,Maris:1997tm}.
The usual argument for a quark vacuum condensate is the Gell-Mann--Oakes--Renner formula:
$
m^2_\pi = -2 m_q {\langle0| \bar q q |0\rangle/ f^2_\pi}.
$
However, in the Bethe-Salpeter and light-front formalisms, where the pion is a $q \bar q$ bound-state, the GMOR relation is replaced by
$
m^2_\pi = - 2 m_q {\langle 0| \bar q \gamma_5  q |\pi \rangle/ f_\pi},
$
where $\rho_\pi \equiv - \langle0| \bar q \gamma_5  q |\pi\rangle$  represents a pion decay constant via an an elementary pseudoscalar current. 

The cosmological constant measures the matrix element of the energy momentum tensor $T^{\mu \nu}$ in the background universe.  It corresponds to the measurement of the gravitational interactions of  a probe of finite mass;   it only senses the causally connected domain within the light-cone of the observer.  If the universe is empty, the appropriate vacuum state is thus the LF vacuum since it is causal.  One automatically obtains a vanishing cosmological constant from the LF vacuum.
Thus, as argued in Refs. ~\cite{Brodsky:2008be,Brodsky:2008xu,Brodsky:2009zd}   the 45 orders of magnitude conflict of QCD with the observed value of the cosmological condensate is removed, and 
a new perspective on the nature of quark and gluon condensates in
QCD  is thus obtained.~\cite{Brodsky:2008be,Brodsky:2008xu,Brodsky:2009zd}.

\begin{acknowledgements}
I am grateful to the organizers of EXA2011 for their invitation to this outstanding interdisciplinary conference.
 I thank all of my collaborators whose work has been cited in this report, particularly Guy de Teramond, Rich Lebed, Ivan Schmidt, and Dae Sung Hwang.
This research was supported by the Department of Energy  contract DE--AC02--76SF00515.  

\end{acknowledgements}

\end{document}